\documentclass[a4paper,12pt,oneside]{article}
\usepackage{jheppub}
\usepackage{microtype}
\allowdisplaybreaks
\hypersetup{
 pdftitle={The LZ inelastic splitting as a probe of anomaly mediation},
 pdfauthor={Seung J. Lee and Taewook Youn},
 pdfsubject={Anomaly mediation, pseudo-Dirac dark matter, and warped sequestering}}
\newcommand{\keV}{\,\mathrm{keV}}\newcommand{\MeV}{\,\mathrm{MeV}}
\newcommand{\GeV}{\,\mathrm{GeV}}\newcommand{\TeV}{\,\mathrm{TeV}}
\newcommand{\st}{\sin^2\!\theta}
\newcommand{\MP}{M_{\rm Pl}}

\title{The LZ inelastic splitting as a probe of anomaly mediation}

\author[a]{Seung J. Lee}
\author[a,b]{Taewook Youn}
\affiliation[a]{School of Physics and Quantum Universe Center, Korea Institute for Advanced Study,\\
85 Hoegi-ro, Dongdaemun-gu, Seoul 02455, Korea}
\affiliation[b]{Gravity, Spacetime, and Particle Physics (GRASP) Initiative, Harvard University,\\
17 Oxford Street, Cambridge, MA 02138, USA}
\emailAdd{sjjlee@kias.re.kr}
\emailAdd{taewookyoun@fas.harvard.edu}

\abstract{
The LZ $248\keV$ recoil candidate admits an interpretation in terms of
pseudo-Dirac dark matter with a mixing-suppressed inelastic $Z$ interaction.
In the MSSM, the Higgsino limit conflicts with the adopted solar-neutrino bound,
while a gaugino realization requires an unprotected mass degeneracy.
We realize such spectra in the MRSSM,
where an $R$-symmetric Dirac bino produces the required $Z$ transition and wino
coannihilation supplies the observed relic abundance. We impose the Higgs and gauge
coupling relations of $\mathcal N=2$ supersymmetry at a high scale and evolve them
to low energy. With the remaining model parameters held fixed, we obtain
spectra spanning roughly $300$--$850\GeV$ that satisfy the adopted constraints
and have a small diagonal axial current. If anomaly mediation dominates the Majorana
splitting, its local scale is tens of MeV for splittings of a few hundred keV.
A five-dimensional warped construction generates this scale through
Goldberger--Wise stabilization and a loop correction to the no-scale K\"ahler potential. The spectra
remain testable through elastic scattering, solar neutrinos and long-lived
charged states.
}

\begin{document}
\maketitle
\raggedbottom

\section{Introduction}
Direct-detection searches usually obtain their greatest sensitivity at recoil energies
of a few tens of keV. The LZ high-energy analysis instead reported one
nuclear-recoil-like event at $248\keV$, with an expected background of $0.01$ events
\cite{LZ:2026axp}. The event has prompted interpretations based on extra dimensions, new
mediators, composite and atomic states, and isotope dependence
\cite{Lee:2026xxh,Cui:2009xq,Ahmed:2026com,Ahmed:2026qjg,Ahmed:2026kan,Jung:2026otm,Sannino:2026hkc,Chanda:2026wdl,Tan:2026ljg,DiMauro:2026syh}.
Electroweak-fermion and supersymmetric realizations include singlet--doublet,
Higgsino, bino, singlino, sneutrino, vector-like-lepton and MRSSM constructions
\cite{Lee:2026jxl,Bisal:2026khf,Khan:2026zuj,Lian:2026hpm,Langhoff:2026qgo,Visinelli:2026kgt,Lee:2026wof,Fan:2026hzw,Yin:2026jnn,Pallis:2026jdc,Roy:2026wzb,Lian:2026bze,Hor:2026voa,Khan:2026pgu,Elahi:2026vlm,Jeong:2026tlf}.
Scalar and portal models provide another class
\cite{Murayama:2026apt,CentellesChulia:2026ats,Jodlowski:2026wjx,De:2026win,Arcadi:2026kev},
while inelastic fermion dark matter has also been linked to Dirac neutrino masses
\cite{Das:2026uyy}. Other proposals invoke a PeV-scale Higgsino, exothermic dark
matter, neutrino-induced scattering, or evaporating primordial black holes
\cite{Unwin:2026qep,Sheikh:2026lxx,Deka:2026zeu,Brdar:2026ukx,Kalita:2026hvb}.
Atmospheric-neutrino up-scattering is constrained in Ref.~\cite{AristizabalSierra:2026vlr}.

Model-independent likelihood analyses and comparisons of boosted and inelastic
interpretations are given in Refs.~\cite{Palmisano:2026kuj,Mahapatra:2026glu}.
Recoil kinematics and neutrino-background limitations for inelastic scattering
are examined in Refs.~\cite{DiMauro:2026ldr,Amarasinghe:2026iai}.
Solar-evolution and gamma-ray studies provide complementary tests
of these scenarios \cite{Qi:2026vyp,Capolupo:2026uyi,Dutra:2026fng,Abazajian:2026zgg}.

In this paper, we focus on
endothermic inelastic dark matter, in which producing the heavier neutral state
consumes part of the incident kinetic energy. The recoil spectrum therefore ties the
dark-matter mass, level splitting and transition strength together. The
singlet--doublet analysis of Ref.~\cite{Lee:2026jxl} shows how mixing can suppress
an inelastic transition through the Standard-Model $Z$.
Supersymmetry controls the radiative sensitivity of scalar masses and provides
a framework for addressing the hierarchy problem
\cite{Witten:1981nf,Dimopoulos:1981zb,Kaul:1981wp,Haber:1984rc}. It also relates the
inelastic splitting to a small source of supersymmetry breaking through anomaly
mediation.

A pseudo-Dirac spectrum is protected by an approximate conserved charge: a Dirac mass
preserves it, while a small Majorana term breaks it and sets the splitting.
Small singlet Majorana masses have been used in pseudo-Dirac singlet--doublet
models of dark matter, neutrino masses and leptogenesis
\cite{Konar:2020wvl,Konar:2020vuu}. In a supersymmetric theory, $U(1)_R$ can play this role. The minimal supersymmetric
standard model (MSSM), however, contains no adjoint
fermion with which the bino can form an $R$-preserving Dirac state. Its available
Higgsino and gaugino limits either conflict with solar capture or reproduce the target
only through a one-part-in-$10^6$ coincidence.

We instead use the minimal
$R$-symmetric supersymmetric standard model (MRSSM) \cite{Kribs:2007ac,Diessner:2014ksa}. Its adjoint
multiplets allow Dirac gaugino masses,
and its additional Higgs doublets complete an $R$-symmetric Higgs sector.
The coupling relations of four-dimensional $\mathcal N=2$ supersymmetry allow
mixing of the bino with the Higgsinos to produce the inelastic transition while
suppressing the diagonal axial $Z$ current. Following the high-scale MRSSM treatment of
Ref.~\cite{Benakli:2018vqz}, we impose these coupling relations at a high scale
and evolve the couplings to low energy with the MRSSM renormalization-group equations.
The same Dirac-gaugino sector contains a
slightly heavier wino-like state; its coannihilation with the bino produces the
observed thermal relic abundance. We scan the dark-matter mass and splitting,
adjusting the bino, wino and down-type Higgsino masses at each point while
holding $\tan\beta$, the high scale and the heavier-particle masses fixed.
For these inputs, spectra satisfying the adopted constraints have dark-matter
masses spanning roughly $300$--$850\GeV$.
We use the $730\GeV$, $300\keV$ point as a benchmark for the detailed calculations.
Elastic scattering, solar neutrinos and charged-wino searches test this broader
family of spectra.
The MRSSM avoids the accidental mass cancellation of the MSSM gaugino realization:
its small splitting is technically natural.

The small Majorana term has a further implication. A nonzero gravitino mass breaks
the continuous $R$ symmetry, and anomaly mediation communicates that breaking to the
gauginos. If it dominates, a $300\keV$ benchmark splitting corresponds to a local
anomaly-mediated scale of about $31\MeV$. In a warped theory this scale is the product of the
gravitino mass and the warp factor, multiplied by a coefficient that depends on how
the no-scale structure is broken. The recoil spectrum therefore probes the combination
$m_{3/2}M_{\rm IR}/\MP$, rather than the gravitino mass alone. Bino stability still
requires $m_{3/2}>m_\chi$, and a thermally produced gravitino motivates the stronger
$60\TeV$ reference.
This is why we combine the MRSSM with warping: an ordinary weak-scale
$\mu H_u\!\cdot H_d$ term breaks $U(1)_R$ and
would generically produce a bino splitting far above $300\keV$.
The $R$-Higgs doublets of the MRSSM avoid this source of breaking,
while an unsuppressed compensator, as in minimal flat-space sequestering,
would require $m_{3/2}\simeq30\MeV$ for a
$300\keV$ splitting, making the bino unstable.

We implement this separation using the five-dimensional matching
of Ref.~\cite{Nee:2025nmi}. Our primary realization stabilizes the radion with a
Goldberger--Wise hypermultiplet and breaks the no-scale structure through a loop
correction to the K\"ahler potential. Supersymmetry is broken by a bulk constant
superpotential; the ultraviolet-brane field supplies only the small positive uplift.
The resulting infrared compensator is warped and loop suppressed, whereas the bulk
radion auxiliary field has no tree-level coupling to the infrared-brane MRSSM. The
$D'$ energy that generates the Dirac masses shifts the radion minimum
negligibly. A bulk-condensate model gives an alternative realization with a lower
infrared scale. Within the sequestered brane action, gravitational loops enter an
infrared-brane gauge kinetic function through the local compensator and do not generate
an independent Majorana term \cite{Randall:1998uk,Luty:2001jh,Luty:2001zv}.

This paper is organized as follows. Sec.~\ref{sec:nogo} explains why the MSSM
does not naturally realize the target. Sec.~\ref{sec:action} then defines the
sequestered MRSSM framework and derives the
anomaly-mediated splitting. Sec.~\ref{sec:embed} presents the neutralino mass matrix
and viable spectra, and Sec.~\ref{sec:pred} gives their low-energy predictions.
Appendix~\ref{sec:seq} gives the matching, sequestering and radion-backreaction checks
for the two warped realizations.

\section{Why the MSSM does not naturally realize the target}\label{sec:nogo}
The target requires more than two neutralinos separated by $300\keV$. They must form an
approximate Dirac pair, so that the $Z$ current is predominantly an off-diagonal vector
transition, and its strength must be reduced to $\st\simeq0.02$ at the $730\GeV$ reference
point. The MSSM offers two possible limits, but each supplies only part of this
structure.

In the basis
$\mathbf n^T=(\tilde B,\tilde W^0,\tilde H_d^0,\tilde H_u^0)$, the neutralino mass
matrix is
\begin{equation}
 \mathcal M_N^{\rm MSSM}=\begin{pmatrix}
 M_1&0&-g'v_d/2&+g'v_u/2\\
 0&M_2&+gv_d/2&-gv_u/2\\
 -g'v_d/2&+gv_d/2&0&-\mu\\
 +g'v_u/2&-gv_u/2&-\mu&0
 \end{pmatrix}.
\label{eq:mssm-matrix}
\end{equation}
Here $\tan\beta=v_u/v_d$. If $N$ diagonalizes this matrix, the neutralino
$Z$ coupling is proportional to \cite{Haber:1984rc}
\begin{equation}
 O''_{ij}=\frac12\left(N_{i3}N^*_{j3}-N_{i4}N^*_{j4}\right).
\label{eq:mssm-z}
\end{equation}
For real parameters one may write $N_{ij}=\eta_iZ_{ij}$, where $Z$ is real and
$\eta_i=1$ or $i$ according to the sign of the corresponding eigenvalue
\cite{Choi:2001ww}. An opposite-sign pair has a coherent vector transition, whereas a
same-sign pair has an axial transition.

When the gauginos are heavy, the two neutral Higgsinos form a pseudo-Dirac pair protected
by approximate Higgsino number. For $M_1=M_2=M\gg|\mu|$, integrating out the gauginos gives
\cite{Nagata:2014wma,Bisal:2026khf}
\begin{equation}
 \delta\simeq\frac{m_Z^2}{M},
 \qquad \delta=300\keV\ \Longrightarrow\ M\simeq3\times10^7\GeV.
\label{eq:higgsino-limit-main}
\end{equation}
A cancellation between the bino and wino contributions can lower the gaugino
scale for non-universal masses \cite{Du:2026guj}. It leaves the Higgsino's
full-strength $Z$ transition unchanged, $\st=1$. At
$730\GeV$ this would give thousands of LZ recoils for a $300\keV$ splitting. Moving
the splitting to the kinematic edge can reduce the xenon rate, but then solar capture
conflicts with the adopted IceCube bound in the thermalized-population treatment
\cite{Lee:2026jxl,Pospelov:2026ewn,Nguyen:2026lui,Bose:2026ndd,Lee:2026xxh}.
The Higgsino limit therefore protects the splitting but not
the required transition strength.

A gaugino-dominated pair can suppress the transition if $M_1\simeq-M_2$, but the
bino and wino soft masses are independent. For $\tan\beta=5$, the numerical solution
lies near
\begin{equation}
 M_1=734\GeV,\qquad M_2=-734\GeV,\qquad \mu=1.02\TeV.
\label{eq:mssm-root}
\end{equation}
Higgsino mixing must cancel nearly all of the $0.07\GeV$ difference between the bino
and wino mass magnitudes. We quantify this with the Barbieri--Giudice sensitivity
\cite{Barbieri:1987fn}, $\Delta_p=|(p/\delta)\,\partial\delta/\partial p|$,
for $p=M_1,M_2,\mu,\tan\beta$, holding the other inputs fixed. It measures the
fractional change in the splitting caused by a fractional change in each input.
Here it quantifies the accidental small splitting.
At this solution,
\begin{equation}
 \Delta_{M_1}\simeq\Delta_{M_2}\simeq2.4\times10^6,\qquad
 \Delta_\mu=2.2\times10^4,\qquad
 \Delta_{\tan\beta}=1.9\times10^4.
\label{eq:mssm-tuning}
\end{equation}
Changing the sign of $\mu$ or varying $2\leq\tan\beta\leq50$ does not remove this
sensitivity. The solutions also have
$|O''_{11}|/|O''_{12}|=0.07$--$0.3$, and hence retain an elastic spin-dependent
$Z$ channel. The gaugino limit can reproduce the desired spectrum, but only through
an unprotected mass degeneracy.

The MRSSM avoids this obstruction by pairing the bino with the fermion in the
hypercharge-adjoint multiplet. The resulting $R$-preserving Dirac mass fixes the
leading spectrum, while a much smaller Majorana term breaks $U(1)_R$ and produces the
inelastic splitting. After this breaking is included, the physical states are two Majorana
eigenstates; ``Dirac bino'' refers to the symmetry limit that protects their
near-degeneracy.

\section{A sequestered MRSSM framework}\label{sec:action}
The construction separates the $R$-preserving TeV spectrum from the much smaller
$R$-breaking insertion. The infrared-boundary action supplies the Dirac masses, while
the warped bulk fixes the local compensator that supplies the Majorana masses. We
describe these two sectors in turn and combine them through anomaly mediation.

\subsection{The visible sector and Dirac masses}\label{sec:visible}
The infrared-boundary theory is the MRSSM \cite{Kribs:2007ac}. It extends the MSSM by an
adjoint chiral multiplet for each gauge group and two
$R$-Higgs doublets. The adjoint fermions provide Dirac partners for the gauginos,
while the additional doublets permit Higgsino masses without breaking the continuous
$U(1)_R$. We use the field and sign conventions of Ref.~\cite{Diessner:2014ksa}. In these
conventions, $U(1)_R$ assigns charge zero to $H_{u,d}$ and to the adjoint chiral
superfields.
In the electroweak sector these are $S$, the $U(1)_Y$ adjoint, and $T$, the
$SU(2)_L$ adjoint. Their fermions, $\widetilde S$ and $\widetilde T$, pair with
the bino and wino; the analogous color adjoint completes the Dirac gluino sector.
The $R$-Higgs fields $R_{u,d}$ have charge two, while the matter
superfields have charge one.

\paragraph{The $\mathcal N=2$ gauge--Higgs action.}
For each gauge factor, an $\mathcal N=2$ vector multiplet consists of an
$\mathcal N=1$ vector $V_i$ and its adjoint chiral partner $\Phi_i$. The two
hypermultiplets are $(H_d,H_d^c)$ and $(H_u,H_u^c)$, where
$H_a^c=R_a^T\epsilon$ is the row superfield in the conjugate gauge representation
and $\epsilon_{12}=1$. With canonical kinetic terms, their complete renormalizable
gauge--Higgs action can be written in $\mathcal N=1$ superspace as
\begin{align}
 \mathcal L_{\mathcal N=2}^{\rm gauge\text{-}Higgs}
 ={}&\int d^4\theta\left[
 \sum_i\Phi_i^\dagger e^{2g_iV_i^{\rm adj}}\Phi_i
 +\sum_{a=u,d}\left(H_a^\dagger e^{2\mathcal V_a}H_a
 +H_a^c e^{-2\mathcal V_a}H_a^{c\dagger}\right)\right]\nonumber\\
 &+\left\{\int d^2\theta\left[
 \frac14\sum_i W_i^\alpha W^i_\alpha+W_{\mathcal N=2}\right]
 +\mathrm{h.c.}\right\},\nonumber\\
 W_{\mathcal N=2}
 ={}&\sum_{a=u,d}H_a^c
 \left(\mu_a+\sqrt2g'Y_aS+\sqrt2g\,T^At^A\right)H_a.
 \label{eq:n2-gauge-higgs-action}
\end{align}
The gauge sum includes $U(1)_Y$, $SU(2)_L$ and $SU(3)_c$, with contractions
over adjoint indices understood; the Higgs hypermultiplets are color singlets.
Here $Y_d=-1/2$, $Y_u=1/2$, $t^A=\sigma^A/2$, and
$\mathcal V_a=g'Y_aV_Y+gV_W^At^A$.
In the conventions of Ref.~\cite{Diessner:2014ksa}, the triplet matrix is
$T=T^A\sigma^A/\sqrt2$. Expanding the last line therefore gives
$\lambda_d=-g'/\sqrt2$, $\lambda_u=g'/\sqrt2$, and
$\Lambda_d=\Lambda_u=g$. Rephasing $R_d\to-R_d$ reverses
$\mu_d$, $\lambda_d$ and $\Lambda_d$ together and gives the signs used below.
The $\mu_a$ terms are supersymmetric hypermultiplet masses and do not themselves
break $\mathcal N=2$. This is the gauge--Higgs sector used in the high-scale
MRSSM analysis of Ref.~\cite{Benakli:2018vqz}.

\paragraph{Breaking to $\mathcal N=1$.}
We couple this sector to three generations of chiral quarks and leptons through
the boundary interaction
\begin{equation}
 \Delta\mathcal L_{\mathcal N=1}^{\rm matter}
 =\int d^4\theta\sum_{X=Q,U^c,D^c,L,E^c}X^\dagger e^{2\mathcal V_X}X
 +\left\{\int d^2\theta\,W_Y+\mathrm{h.c.}\right\}.
 \label{eq:n2-to-n1-deformation}
\end{equation}
Here $\mathcal V_X$ is the gauge connection in the representation of $X$, and
$W_Y$ is the quark and lepton Yukawa superpotential in the first line of
Eq.~(\ref{eq:mrssm-W}). These matter multiplets have no conjugate
hypermultiplet partners. Their gauge interactions and $W_Y$ preserve the first
supersymmetry but explicitly break the second. Thus the
$\mathcal N=2\to\mathcal N=1$ step is specified as a boundary interaction,
not as spontaneous breaking of a complete $\mathcal N=2$ matter theory.
The surviving $U(1)_R$ admits the charge assignment stated above.

At the matching scale, combining these terms and the supersymmetry-breaking
operators gives the visible-sector action
\begin{align}
 \mathcal L_{\rm vis}={}&\int d^4\theta\sum_X X^\dagger e^{2\mathcal V_X}X
 +\left\{\int d^2\theta\left[
 \frac14\sum_i W_i^\alpha W^i_\alpha+W_{\rm MRSSM}\right]
 +\mathrm{h.c.}\right\}\nonumber\\
 &+\left\{\int d^2\theta\,
 \frac{\sqrt2}{M_*}\sum_i c_i^D W'_\alpha W_i^\alpha\Phi_i
 +\mathrm{h.c.}\right\}-V_{\rm soft},
 \label{eq:full-mrssm-action}\\
 V_{\rm soft}={}&\sum_X m_X^2|X|^2+
 \left(B_\mu H_u\!\cdot H_d+\mathrm{h.c.}\right).
 \label{eq:mrssm-soft}
\end{align}
Here $X$ runs over the Higgs, $R$-Higgs, adjoint and matter multiplets, with
$\mathcal V_X$ in the corresponding representation. The supersoft operator is
suppressed by the messenger scale $M_*$ and has dimensionless coefficients $c_i^D$.
The $R$ symmetry forbids
Majorana gaugino masses, the MSSM $\mu H_u\!\cdot H_d$ term and trilinear $A$-terms.
The superpotential is
\begin{equation}
\begin{aligned}
 W_{\rm MRSSM}={}&Y_uU^cQ\cdot H_u-Y_dD^cQ\cdot H_d-Y_eE^cL\cdot H_d\\
 &+\mu_dR_d\cdot H_d+\mu_uR_u\cdot H_u
 +\lambda_dS R_d\cdot H_d+\lambda_uS R_u\cdot H_u\\
 &+\Lambda_dR_d\cdot T H_d+\Lambda_uR_u\cdot T H_u.
\end{aligned}
\label{eq:mrssm-W}
\end{equation}
Eq.~(\ref{eq:n2-gauge-higgs-action}) fixes the gauge--Higgs couplings before
the $\mathcal N=1$ deformation. We take its tree-level relations as the
boundary condition at the infrared-brane cutoff $M_{\rm IR}$, in our
rephased convention:
\begin{equation*}
 \lambda_u(M_{\rm IR})=\lambda_d(M_{\rm IR})
 =\frac{g'(M_{\rm IR})}{\sqrt2},\qquad
 \Lambda_u(M_{\rm IR})=-\Lambda_d(M_{\rm IR})=g(M_{\rm IR}).
\end{equation*}
The warped construction below determines the supersymmetry-breaking scales;
the gauge--Higgs matching condition is an assumption about the brane theory at
its cutoff. We retain the adjoints and $R$-Higgs fields as independent brane
multiplets rather than obtain them by projecting a bulk gauge multiplet.

Below $M_{\rm IR}$ we evolve the four couplings with the two-loop MRSSM equations
rather than imposing these equalities again at low energy. For
$M_{\rm IR}=10^{14}\GeV$, the ratios
$\sqrt2\lambda_d/g'$, $\sqrt2\lambda_u/g'$, $-\Lambda_d/g$ and $\Lambda_u/g$
at $10\TeV$ are approximately $1.10$, $0.99$, $1.33$ and $1.21$ at the retuned
reference point. Finite corrections to the high-scale matching are not included.
Related high-scale $\mathcal N=2$ boundary conditions and their evolution, together
with anomaly mediation, were studied in Ref.~\cite{Yin:2016pkz}.
Sec.~\ref{sec:embed} uses these evolved couplings throughout the numerical
calculation.

The hidden Abelian field strength $W'_\alpha$ is treated as a background source
for the visible action. Its
$D'$ expectation value in the supersoft term of
Eq.~(\ref{eq:full-mrssm-action}) supplies the $R$-preserving Dirac masses \cite{Fox:2002bu}:
\begin{equation}
 \langle W'_\alpha\rangle=\theta_\alpha D',\qquad
 M_i^D\propto c_i^D\frac{D'}{M_*}.
\label{eq:supersoft-review}
\end{equation}
This separates the two ingredients of the construction: $D'$ generates the TeV
Dirac spectrum without breaking $R$, while the much smaller Majorana insertion
must arise from a distinct $R$-breaking source. Eqs.~(\ref{eq:full-mrssm-action})--
(\ref{eq:supersoft-review}) therefore define the $R$-symmetric part of the spectrum.
The $D'$ expectation value and $V_{\rm soft}$ break the remaining
$\mathcal N=1$ supersymmetry; anomaly mediation subsequently adds the small
$R$-breaking terms.
We next embed this action on the infrared boundary and determine the local
$R$-breaking scale.

\subsection{Warped stabilization and the local compensator}\label{sec:warped-setup}
\begin{figure}[t]
\centering
\includegraphics[width=0.90\textwidth]{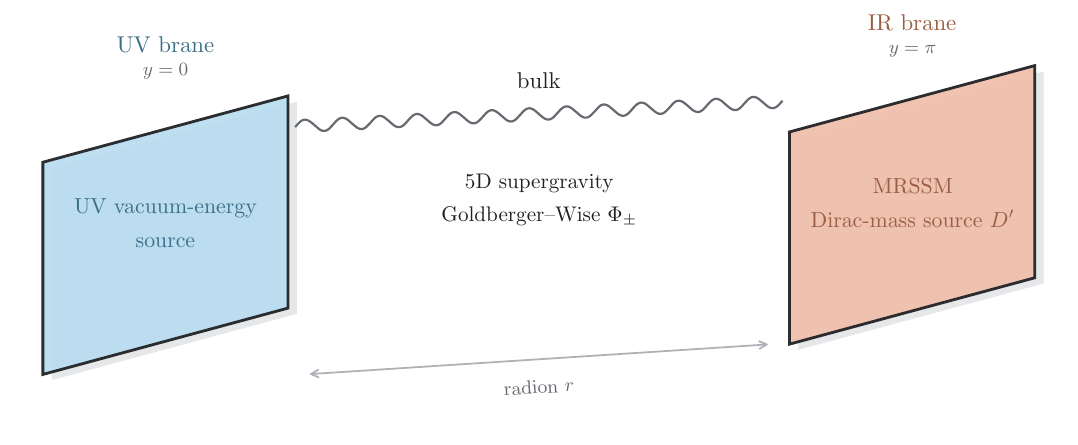}
\caption{Sector localization in the warped construction of
Eq.~(\ref{eq:matched-5d-action}). The ultraviolet brane carries a vacuum-energy
source, while the infrared brane carries the MRSSM and its supersoft source $D'$.
Five-dimensional supergravity and the Goldberger--Wise hypermultiplet $\Phi_\pm$
propagate in the bulk.}
\label{fig:warped-branes}
\end{figure}

The five-dimensional sector determines the local compensator scale $F_{\rm IR}$.
We place the MRSSM action of
Eq.~(\ref{eq:full-mrssm-action}), including its $D'$ source, on the infrared boundary
of a supersymmetric warped background of Randall--Sundrum type. The metric, warp
factor and infrared scale are
\begin{equation}
 ds_5^2=e^{-2kry}g_{\mu\nu}dx^\mu dx^\nu-r^2dy^2,\qquad
 \rho=e^{-k\pi r},\qquad M_{\rm IR}=k\rho,\qquad \MP^2\simeq\frac{M_5^3}{k}.
\label{eq:warped-geometry}
\end{equation}
Here $k$ is the curvature scale, $r$ is the interval modulus, $M_5$ is the
five-dimensional Planck scale, and $\MP$ is the reduced four-dimensional Planck
mass; the last relation assumes $\rho\ll1$. Fig.~\ref{fig:warped-branes}
summarizes the field localization.
The interval boundary conditions preserve four of the bulk's eight supercharges,
so gravity couples to the boundary through $\mathcal N=1$ multiplets. This
projection concerns the bulk fields; it does not project out the independent
brane adjoints in Eq.~(\ref{eq:n2-gauge-higgs-action}).
Here $\sigma(y)=-kry$. Ref.~\cite{Nee:2025nmi} recently derived the auxiliary fields
directly in five dimensions and showed that the older four-dimensional theory with
independent constant superpotentials on the two boundaries does not reproduce the
five-dimensional potential. We therefore use that matched five-dimensional framework,
rather than that boundary-superpotential effective theory. Our application adds the
infrared MRSSM sector, fixes the compensator scale from the inelastic splitting, and
tests the resulting spectrum and backreaction. The phenomenological tests can be
performed independently of the radion-stabilization calculation.

For the primary realization, the terms relevant to stabilization and mediation are
\cite{Abe:2004ar,PaccettiCorreia:2006rpo}
\begin{align}
 S_5={}&\int d^4x\sqrt{-g}\int_0^\pi dy\,
 \bigl[\mathcal L_{\rm SUGRA}^{(5)}(C_\pm,\mathcal V_\Sigma)
 +\mathcal L_{\rm hyp}^{(5)}(\Phi_\pm;\mathcal V_\Sigma)
 +\mathcal L_J+\Delta\mathcal L_\beta
 +\mathcal L_{W_{\rm bulk}}\bigr]\nonumber\\
 &+\int_{y=0}d^4x\sqrt{-g_0}\,\mathcal L_{\rm uplift}
 +\int_{y=\pi}d^4x\sqrt{-g_\pi}\,\mathcal L_{\rm vis},\nonumber\\
 \mathcal L_J={}&4e^{3\sigma}\left[
 -\delta(y)\!\int d^2\theta\,C_+J_0\Phi_+
 +\delta(y-\pi)\!\int d^2\theta\,C_+J_\pi\Phi_+
 +{\rm h.c.}\right],\nonumber\\
 \Delta\mathcal L_\beta&\supset
 e^{2\sigma}\!\int d^4\theta\,\frac{\beta}{3\mathcal V_\Sigma^3}
 |C_+|^{4/3},\nonumber\\
 \mathcal L_{W_{\rm bulk}}&=-M_5r e^{3\sigma}
 \left[\int d^2\theta\,W_{\rm bulk}(C_+^2+C_-^2)+{\rm h.c.}\right].
\label{eq:matched-5d-action}
\end{align}
The bulk superspace densities include the warp and radius factors explicitly;
$g$ is the four-dimensional metric in Eq.~(\ref{eq:warped-geometry}).
The physical gravity multiplet contains the five-dimensional metric, the
graviphoton and two gravitini. In the four-dimensional superfield description,
$\Sigma$ contains the radius and the fifth component of the graviphoton, while
$V$ contains its four-dimensional vector component. Their gauge-invariant
combination is $\mathcal V_\Sigma=\Sigma+\Sigma^\dagger-\partial_y V$.
The compensator superfields $C_\pm$ implement the off-shell supergravity
description; they are not additional physical matter multiplets \cite{Nee:2025nmi}.

The hypermultiplet $\Phi_\pm$ and its boundary sources $J_0,J_\pi$ stabilize the
radius. The constant $W_{\rm bulk}$ breaks supersymmetry, and the loop coefficient
$\beta\sim1/(16\pi^2)$ breaks the no-scale K\"ahler structure.

The term $\mathcal L_{\rm uplift}$ supplies positive vacuum energy on the
ultraviolet boundary, adjusted to make the total four-dimensional vacuum energy
approximately zero. Appendix~\ref{sec:bulk-fields} gives the bulk densities and
the ultraviolet boundary action, while the component reduction
follows in Appendix~\ref{sec:gw-reduction}.

\paragraph{Goldberger--Wise stabilization with loop corrections.}
Writing the odd scalar as $\varphi_-=\Theta(y)\widetilde\varphi_-$, the two source
terms in Eq.~(\ref{eq:matched-5d-action}) impose
\begin{equation*}
 \widetilde\varphi_-(0)=M_5^{3/2}J_0,\qquad
 \widetilde\varphi_-(\pi)=M_5^{3/2}J_\pi.
\end{equation*}
The bulk $F$-flat solution satisfies both conditions only when
$\rho_0^\epsilon=J_\pi/J_0$. Away from this value, substituting the bulk profile back
into the positive $F$-term action gives
\begin{equation}
 V_{\rm GW}=(4+2\epsilon)kM_5^3\rho^{4+2\epsilon}
 \left|J_0-J_\pi\rho^{-\epsilon}\right|^2+O(\rho^8),
 \qquad \rho_0^\epsilon=\frac{J_\pi}{J_0}.
\label{eq:gw-potential}
\end{equation}
The appendix displays the bulk profile and the integration leading to
Eq.~(\ref{eq:gw-potential}). Its minimum is supersymmetric and fixes the radius before
the much smaller no-scale-breaking perturbation is included.

\paragraph{Generating the infrared compensator.}
With the radius fixed, varying Eq.~(\ref{eq:matched-5d-action}) with respect to the
radion and compensator auxiliaries, and retaining the first nonzero order in $\beta$,
gives
\begin{equation}
 F_\Sigma=-M_5^2rW_{\rm bulk}e^\sigma+O(\beta),\qquad
 F_{C_+}=-\frac{2\beta W_{\rm bulk}}{M_5^{3/2}r^4}e^\sigma+O(\beta^2).
\label{eq:matched-auxiliaries}
\end{equation}
The first relation identifies
$m_{3/2}=|F_\Sigma(0)|/(M_5r)=M_5|W_{\rm bulk}|$ at the ultraviolet boundary,
in the no-scale convention \cite{Cremmer:1983bf,Ellis:1984kd,Lahanas:1986uc}. Dividing the
second relation by $C_+=M_5^{3/2}$, evaluating $e^\sigma$ at $y=\pi$, and substituting
$W_{\rm bulk}=m_{3/2}/M_5$ gives
\begin{equation}
 F_{\rm IR}\equiv\left|\frac{F_{C_+}}{C_+}\right|_{y=\pi}
 =c_\beta\rho_0m_{3/2},\qquad
 c_\beta=\frac{2\beta}{(M_5r)^4},\qquad
 M_5r=\frac{\ln(1/\rho_0)}{\pi(k/M_5)}.
\label{eq:matched-loop-scale}
\end{equation}
In Eq.~(\ref{eq:matched-loop-scale}), $W_{\rm bulk}$ sets $m_{3/2}$, the stabilized radius sets
$\rho_0$, and the loop and volume factors in $c_\beta$ suppress the local
$R$-breaking scale. Imposing $F_{\rm IR}\simeq31\MeV$, the value required by the
$300\keV$ splitting derived in Sec.~\ref{sec:measure}, and taking
$\beta=(1$--$4)/(16\pi^2)$, $0.3\leq k/M_5\leq1$, and
$10^4\leq m_{3/2}/\GeV\leq10^7$ gives
$4\times10^{-6}\lesssim c_\beta\lesssim2.5\times10^{-3}$; for $k=M_5$ the range
narrows to $10^{-4}$--$10^{-3}$. The normalization alternatives and the resulting
envelopes in Fig.~\ref{fig:matched-mediation} are detailed in
Appendix~\ref{sec:seq}.

\paragraph{Coupling to the infrared theory.}
The radion auxiliary field $F_\Sigma$ does not couple at tree level to
boundary-localized matter, and boundary pieces of $F_{C_+}$ are localized on the brane
carrying the corresponding superpotential. With $W_\pi=0$ on the
$R$-symmetric infrared brane, the MRSSM therefore sees the warped bulk contribution
$F_{C_+}(\pi)$. Gravitational corrections to a boundary gaugino are likewise encoded
in this local compensator, as discussed below. For an unsuppressed supersoft
operator, the $D'$ energy shifts the Goldberger--Wise minimum by less than
$7\times10^{-12}$, so its effect on the stabilized radius is negligible.

\paragraph{Stabilization by a bulk gaugino condensate.}
An alternative to Eqs.~(\ref{eq:gw-potential})--(\ref{eq:matched-loop-scale}) replaces
both the Goldberger--Wise stabilizer and the $\beta$ correction by a radius-dependent bulk gaugino condensate
$W_\lambda e^{-2\alpha\Sigma}$; it is not an additional contribution to
Eq.~(\ref{eq:matched-auxiliaries}). Let $p=\alpha W_\lambda$ and
$t=\alpha M_5r$. The stationary condition and the matched constant term give
\begin{equation*}
 p e^{-t}=\frac{3|W_{\rm bulk}|}{2+6/t+3/t^2},\qquad
 m_{3/2}=\frac{M_5|W_{\rm bulk}|}{4}.
\end{equation*}
Substituting these relations into the compensator auxiliary field gives
\begin{equation}
 F_{\rm IR}=M_5p\rho_0e^{-t}=c_1\rho_0m_{3/2},\qquad
 c_1=\frac{12}{2+6/t+3/t^2}.
\label{eq:matched-condensate-scale}
\end{equation}
At the particular vacua we evaluated, $c_1$ is approximately $5.4$--$5.5$.
In the mathematical limit $t\to\infty$, $c_1\to6$.
Requiring a stable bino and a radion below the Kaluza--Klein scale selects
$m_{3/2}=2$--$25\TeV$ and
$M_{\rm IR}=6\times10^{11}$--$7\times10^{12}\GeV$. These gravitino masses lie
below the $60\TeV$ nucleosynthesis reference, so a thermal gravitino population
requires a sufficiently low reheating temperature. Moreover, the uniform bulk condensate of the matched theory has not been
derived from the zero-mode $SU(2)$ stabilizer of Ref.~\cite{Luty:2002ff} with
ultraviolet-brane flavours. We therefore treat this model as an alternative rather
than the primary realization.

\begin{figure}[t]
\centering
\includegraphics[width=0.96\textwidth]{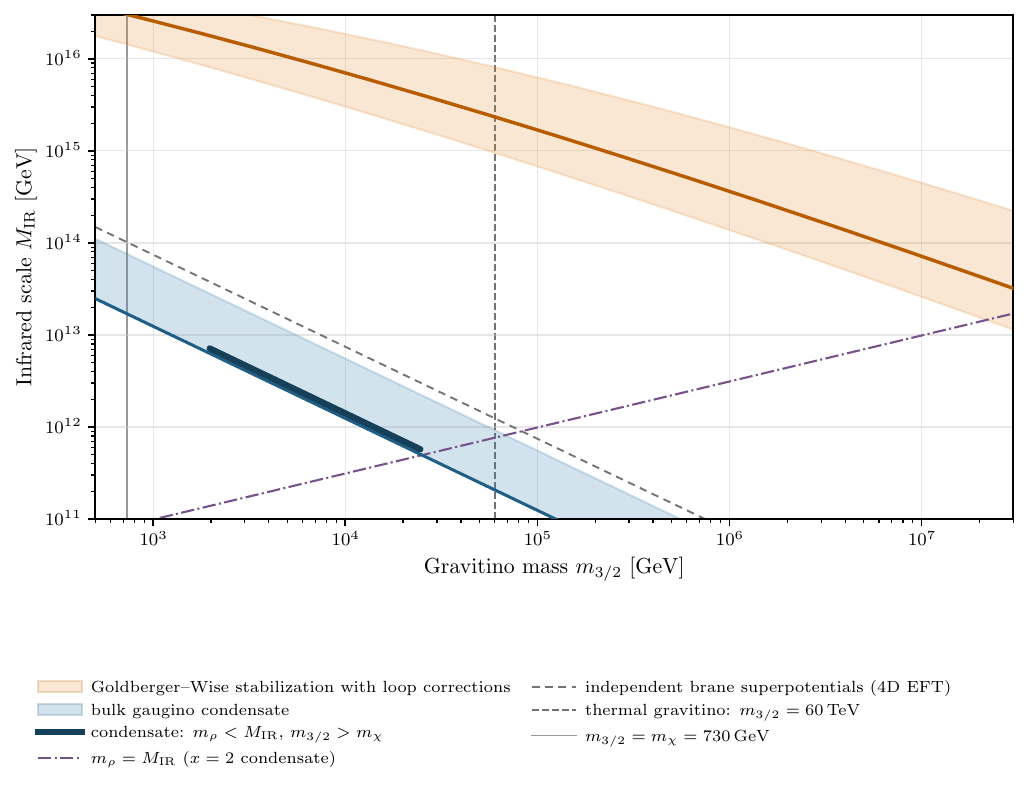}
\caption{Relations between $m_{3/2}$ and $M_{\rm IR}$ for a $300\keV$
splitting. The orange curve uses Goldberger--Wise stabilization with
$\beta=1/(16\pi^2)$ and $k=M_5$; its envelope varies $\beta$, $k/M_5$ and the
$\sqrt3$ normalization. The blue band is the condensate alternative, including its
factor-four gravitino convention; the thick segment obeys both bino stability and
$m_\rho<M_{\rm IR}$. The dot-dashed curve marks $m_\rho=M_{\rm IR}$, while the grey
dashed curve shows the earlier four-dimensional theory with independent boundary superpotentials
\cite{Luty:2000ec,Luty:2002ff}. The vertical lines mark $m_{3/2}=m_\chi=730\GeV$, below which
the bino is unstable, and the $60\TeV$ nucleosynthesis reference for a standard
thermal gravitino history.}
\label{fig:matched-mediation}
\end{figure}

\subsection{The anomaly-mediated splitting}\label{sec:measure}
The auxiliary fields calculated in Sec.~\ref{sec:warped-setup} determine the
infrared value of $F_{C_+}/C_+$, whose magnitude is $F_{\rm IR}$. This quantity
generates Majorana gaugino masses through the conformal anomaly. Although the
classical gauge kinetic term is scale invariant, quantum running introduces a
logarithmic dependence on the local scale and hence on the compensator
\cite{Randall:1998uk,DEramo:2012vvz}. Let $g_i$ be the coupling of gauge group $i$ and
$\beta_{g_i}=d g_i/d\ln\mu$ its beta function, where $\mu$ is the renormalization
scale. The one-loop beta function is
\begin{equation*}
 \beta_{g_i}=\frac{b_i g_i^3}{16\pi^2},\qquad b_i=T_{R_i}-3T_{G_i}.
\end{equation*}
Here $T_{R_i}$ sums the Dynkin indices of the chiral multiplets, and $T_{G_i}$
is the adjoint index.
For canonically normalized gauginos $\lambda_i$, the matched five-dimensional
boundary operator and the resulting Majorana masses are \cite{Nee:2025nmi}
\begin{equation}
 \begin{aligned}
 \Delta\mathcal L_R&=-\frac12\sum_i\frac{\beta_{g_i}}{g_i}
 \left.\frac{F_{C_+}}{C_+}\right|_{y=\pi}\lambda_i\lambda_i+
 \mathrm{h.c.},\\
 m_i^{\rm Maj}&=\frac{\beta_{g_i}}{g_i}F_{\rm IR}
 =\frac{b_i g_i^2}{16\pi^2}F_{\rm IR}.
 \end{aligned}
\label{eq:dtt}
\end{equation}
The second line follows by comparing the first with the canonical mass term
$-\tfrac12\sum_i m_i^{\rm Maj}\lambda_i\lambda_i+\mathrm{h.c.}$.
The common insertion is taken real and positive below. We retain the
bulk-compensator normalization of
Eq.~(\ref{eq:matched-loop-scale}); its relation to the weight-one boundary
compensator is given in Appendix~\ref{sec:seq}. The Majorana masses are therefore
suppressed relative to $F_{\rm IR}$ by a loop factor, while the supersoft Dirac
masses remain set independently by $D'/M_*$ and preserve $U(1)_R$ \cite{Fox:2002bu}.

Anomaly-induced Majorana masses in Dirac-gaugino and pseudo-Dirac dark-matter models
have been discussed in Refs.~\cite{Kribs:2007ac,Kribs:2010md,Hsieh:2007wq,DeSimone:2010tf,Kumar:2011np}.
For hypercharge, the MSSM chiral fields give $b_Y=11$ in the $g'$ normalization and
the two $R$-Higgs doublets add one, so $b_Y=12$. The singlet and triplet adjoints do
not carry hypercharge. For $SU(2)_L$, the triplet and $R$-Higgs fields give $b_2=4$.
Let $f_B$ and $f_W$ be the bino and wino fractions in the $R=+1$ Weyl component
of the light Dirac state. At first order in the Majorana masses,
\begin{equation*}
 m_{\rm light}^{\rm Maj}=f_Bm_B^{\rm Maj}+f_Wm_W^{\rm Maj},\qquad
 \mathcal M_{\rm light}=\begin{pmatrix}
 m_{\rm light}^{\rm Maj}&m_\chi\\ m_\chi&0
 \end{pmatrix}.
\end{equation*}
The two masses are $m_{\chi_{1,2}}\simeq m_\chi\mp m_{\rm light}^{\rm Maj}/2$,
so their splitting is
\begin{equation}
 \begin{aligned}
 m_B^{\rm Maj}&=\kappa_1F_{\rm IR},
 &\kappa_1&=\frac{b_Yg'^2(1\TeV)}{16\pi^2}\simeq0.01,\\
 \delta&\simeq(\kappa_1f_B+\kappa_2f_W)F_{\rm IR},
 &\kappa_2&=\frac{4g^2(1\TeV)}{16\pi^2}.
 \end{aligned}
\label{eq:measure}
\end{equation}
The couplings at $1\TeV$ are obtained by continuing the two-loop MRSSM running
in the $\overline{\rm DR}$ scheme; finite heavy-threshold matching is not included.
Gravitational loops contribute to the bino Majorana mass only through the local
compensator and are part of the one-loop $F_{C_+}(\pi)$ used here; no separate term
arises for an infrared-brane gaugino
\cite{Randall:1998uk,Luty:2001jh,Luty:2001zv,Gherghetta:2001sa,Rattazzi:2003rj,Gregoire:2004nn,Nee:2025nmi}.
At the reference point, $f_B\simeq0.98$ and the wino contribution to the splitting
is below $10^{-4}$ of the bino contribution. We retain both insertions in the full
neutralino calculation. A $300\keV$ splitting requires $F_{\rm IR}\simeq31\MeV$.
With Goldberger--Wise stabilization and loop corrections, this
constrains $c_\beta\rho_0m_{3/2}$, not $m_{3/2}$ alone. The interpretation assumes
sequestering and the absence of an additional local Majorana source. The TeV Dirac
masses remain independent of this small $R$-breaking scale. The induced wino
Majorana mass is also sub-MeV; the gluino contribution vanishes at one loop and the
other anomaly-mediated soft terms are negligible \cite{Kribs:2007ac,Kribs:2010md}.

We assume that the full action and its vacuum preserve $R$-parity.
The anomaly-mediated Majorana masses break continuous $U(1)_R$ but preserve this
discrete symmetry.
The lightest parity-odd particle is then stable. Bino dark matter requires
$m_{3/2}>m_\chi$; a lighter gravitino would allow the bino to decay into it and a
photon. For a standard thermal gravitino history, we use
$m_{3/2}\gtrsim60\TeV$ as the nucleosynthesis reference and require
$T_R\lesssim3\times10^8\GeV$ to keep gravitino decays from appreciably changing the
thermal relic abundance \cite{Gherghetta:1999sw,Kawasaki:2008qe,Bolz:2000fu,Pradler:2006qh}.
Appendix~\ref{sec:gravitino-history} gives this reheating bound.

\section{Neutralino spectra}\label{sec:embed}
\subsection{Mass matrix and mixing}
Having separated the mediation problem from the low-energy spectrum, we now
diagonalize the electroweak fermion sector. Let
$\mathbf n_+^T=(\tilde B,\tilde W^0,\tilde R_d^0,\tilde R_u^0)$ and
$\mathbf n_-^T=(\tilde S,\tilde T^0,\tilde H_d^0,\tilde H_u^0)$ carry fermion
$R$ charges $+1$ and $-1$. With zero adjoint scalar vacuum values, the neutral
mass terms are
\begin{equation}
 -\mathcal L_{\rm neut}=\mathbf n_+^T m_\chi\mathbf n_-
 +\tfrac12m_B^{\rm Maj}\tilde B\tilde B
 +\tfrac12m_W^{\rm Maj}\tilde W^0\tilde W^0+\mathrm{h.c.}.
\label{eq:mrssm-full-lag}
\end{equation}
The Dirac mass matrix is \cite{Diessner:2014ksa}
\begin{equation}
 m_\chi=\begin{pmatrix}
 M_B^D & 0 & -g'v_d/2 & +g'v_u/2\\
 0 & M_W^D & +gv_d/2 & -gv_u/2\\
 -\lambda_dv_d/\sqrt2 & -\Lambda_dv_d/2 & -\mu_d & 0\\
 +\lambda_uv_u/\sqrt2 & -\Lambda_uv_u/2 & 0 & +\mu_u
 \end{pmatrix}.
\label{eq:mrssm-matrix}
\end{equation}
Nonzero adjoint scalar vacuum values shift the $\mu_{d,u}$ entries \cite{Diessner:2014ksa}.
The supersoft $M_B^D$ pairs the bino with $\tilde S$, while $\mu_d$ pairs the
down-type Higgsino with $\tilde R_d^0$; their $v_d$-dependent entries mix the
pairs. Singular-value decomposition gives the Dirac masses. The unitary matrices
$N_1$ and $N_2$ rotate $\mathbf n_+$ and $\mathbf n_-$ to the mass basis and satisfy
\begin{equation}
 N_1^*m_\chi N_2^\dagger=\operatorname{diag}(m_{\chi_1},\ldots,m_{\chi_4}),
\label{eq:neutralino-rotations}
\end{equation}
with the masses ordered increasingly.
All parameters are real in our calculation. The small Majorana terms break
$R$ symmetry and split the light state. We retain all neutralino components in
the calculation.

The bino's vector and axial $Z$ currents depend on the Higgsino content of its
left- and right-handed components. In the reduced matrix containing only the
bino and down-type Higgsinos, $\lambda_d=g'/\sqrt2$ makes the two mixing entries
equal and removes the diagonal axial current. This gives an approximation for the leading vector current
when the wino and up-type Higgsino admixtures are small; it is not used for the relic density or elastic rates.
The nearby wino contributes to coannihilation, and the up-type Higgsinos
contribute to Higgs-mediated elastic scattering. The full matrix and
renormalization-group running need not preserve the axial cancellation.
We define the effective mixing strength $\st$
from the vector $Z$ current in the Dirac limit, normalized to one for a pure
down-type doublet:
\begin{equation}
 \st=\frac12\left(f_{Ld}+f_{Rd}-f_{Lu}-f_{Ru}\right).
\label{eq:transition-strength}
\end{equation}
Here $f_{Ld,u}$ and $f_{Rd,u}$ are composition fractions, with
$f_{Ld}=|(N_1)_{13}|^2$, $f_{Lu}=|(N_1)_{14}|^2$,
$f_{Rd}=|(N_2)_{13}|^2$ and $f_{Ru}=|(N_2)_{14}|^2$.
The $L$ fractions measure the $\tilde R_d^0$ and $\tilde R_u^0$ content of the
left-handed component; the $R$ fractions measure the $\tilde H_d^0$ and
$\tilde H_u^0$ content of the right-handed component. The small Majorana terms
turn this vector current into the leading inelastic transition. At fixed mass
and splitting, its $Z$-exchange rate is proportional to $(\st)^2$.
Thus $\st$ denotes an effective coupling strength, not a single angle that diagonalizes
the full matrix. The rates are evaluated with the physical split states.
At the $730\GeV$ reference point $\st\simeq0.02$. The small up-type components
reduce the transition by about $1\%$ relative to the average down-type fraction.
The axial coefficient is
$a_Z=[(f_{Ld}-f_{Rd})-(f_{Lu}-f_{Ru})]/2$; its physical constraint is the
spin-dependent scattering bound, not an exact-zero condition. At the
point with evolved couplings, $a_Z\simeq-3\times10^{-5}$ remains well below that bound.
This suppression does not require exact $\mathcal N=2$ relations:
near-degeneracy suppresses the down-type contribution by the mass difference
between the bino and Higgsino relative to the Dirac mass. At fixed $\st=0.02$ in the $730\GeV$
spectrum, varying $\sqrt2\lambda_d/g'$ between $0.5$ and $3$ keeps $|a_Z|$
of order $10^{-4}$ or less, and the $Z$-exchange spin-dependent cross section
more than three orders of magnitude below the adopted LZ limit.

In the tree-level submatrix for the bino and down-type Higgsino, enhanced mixing
has $\mu_d<0<M_B^D$ in the conventions of Ref.~\cite{Diessner:2014ksa}. Its two diagonal
entries have the same sign, and the mixing grows as their magnitudes approach
one another. The other relative-sign choice gives a mass sum rather than a
mass difference in the mixing denominator. This sign choice is not fixed by
$\mathcal N=2$. We keep $\mu_d$ signed throughout; the gaps quoted below are
differences of physical pole masses, not of the running input parameters.

For the solutions relevant to LZ, a mass difference of order $10\GeV$ between
the down-type Higgsino and bino gives the required transition strength.
At the benchmark mass, Higgsino coannihilation alone does not reproduce the
observed relic abundance. Coannihilation with a nearby wino is also required;
its Dirac mass parameter has the opposite sign to that of the bino.

\subsection{Viable spectra}\label{sec:benchmark}
We scan the dark-matter mass and splitting, $(m_{\chi_1},\delta)$, for
bino-dominated dark matter, using $\st<0.05$ as an approximate $95\%$ bino-content selection.
This defines the sample, not an experimental exclusion.
At each point, normalization to one expected LZ event fixes the mixing strength.
We adjust the signed bino Dirac mass $M_B^D$ and down-type Higgsino mass $\mu_d$
to obtain that mass and mixing, and the signed wino Dirac mass $M_W^D$ to obtain
the observed relic density, $\Omega h^2=0.12$. The common local compensator scale
$F_{\rm IR}$ sets the Majorana terms and hence the splitting.
We specify the running parameters at $Q=10\TeV$, near the squark mass scale.
We select the recoil interval in which the $55\keV$ threshold is kinematically closed but
the $248\keV$ recoil remains accessible. This selection uses natural xenon
isotopes and the annual halo-speed ceiling of $809\,{\rm km\,s^{-1}}$ in the
rate calculation of Ref.~\cite{Lee:2026jxl}.

\begin{figure}[t]
\centering
\includegraphics[width=\textwidth]{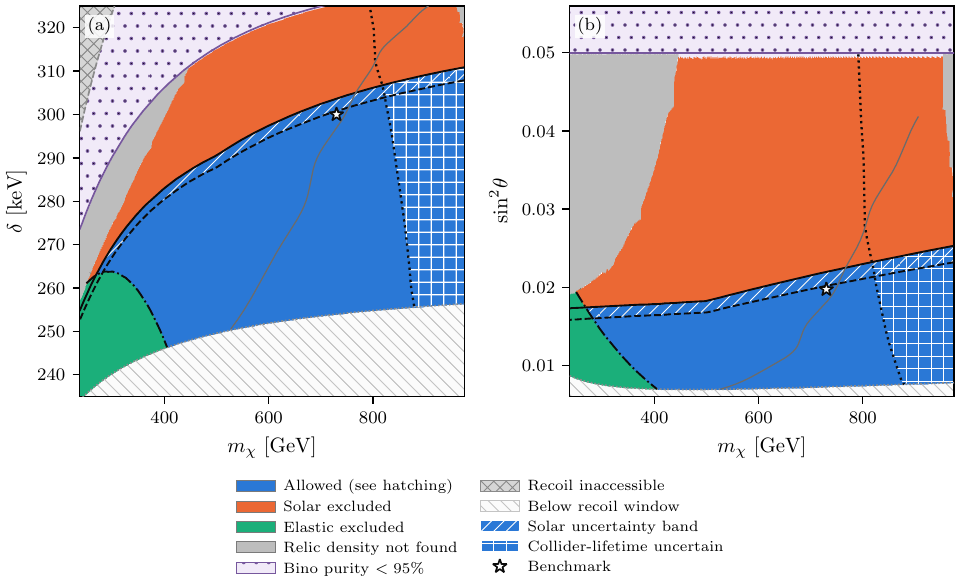}
\caption{MRSSM parameter regions from a scan over $(m_\chi,\delta)$ in (a),
with the corresponding effective mixing strength shown in (b). At each point,
$M_B^D$, $M_W^D$ and $\mu_d$ are adjusted for the mass, one LZ event and
$\Omega h^2=0.12$; $F_{\rm IR}$ sets the splitting and $B_\mu$ keeps
$m_A\simeq3\TeV$. Fixed inputs are $\tan\beta\simeq31$,
$M_{\rm IR}=10^{14}\GeV$, $\mu_u=-2\TeV$ and $M_O^D=3\TeV$.
Scalar mass inputs are $20\TeV$ for electroweak adjoints and $R$-Higgs fields,
$5\TeV$ for the color adjoint, $10\TeV$ for squarks, and $3\TeV$ ($2\TeV$)
for right-handed (left-handed) sleptons. Couplings obey the $\mathcal N=2$
relations at $M_{\rm IR}$ and are evolved to low energy.
Shading interpolates between calculated points.
Unhatched blue passes the adopted tests. Diagonal hatching shows solar
uncertainty; cross hatching shows unresolved collider lifetimes. The solid solar
curve is the nominal bound and the dashed curve includes the tested model
uncertainties, both using the cooling prescription of Ref.~\cite{Lee:2026jxl}.
Solid grey marks relic density not found. Purple stippling lies outside the bino-purity
selection; light diagonal hatching lies below the adopted recoil interval.
Thin grey marks the switch in neutral wino and Higgsino mass ordering.
Thin white edges are unclassified. The star marks the
$730\GeV$, $300\keV$ benchmark.}
\label{fig:mass-scan}
\end{figure}

For Fig.~\ref{fig:mass-scan}, we fix the Higgs-vev ratio
$\tan\beta=v_u/v_d\simeq31$, the high scale $M_{\rm IR}=10^{14}\GeV$ and
the heavier-particle masses at their reference values: $\mu_u=-2\TeV$, $m_A=3\TeV$,
electroweak adjoint and $R$-Higgs scalar masses of $20\TeV$, squark masses of
$10\TeV$, and right- and left-handed slepton masses of $3\TeV$ and $2\TeV$.
The Dirac gluino mass input is $M_O^D=3\TeV$ and the color-adjoint scalar
mass input is $5\TeV$.
The Higgs bilinear $B_\mu$ is adjusted to keep $m_A$ fixed.

At $M_{\rm IR}$ we set the Higgs and gauge couplings to the relations required by
$\mathcal N=2$ supersymmetry, then evolve them to $Q$ using the two-loop equations
of Sec.~\ref{sec:visible}. We repeat this evolution together with the spectrum
calculation until the final couplings satisfy those high-scale relations.
The chosen $\tan\beta$, $M_{\rm IR}$ and heavy-sector values are benchmark inputs, not predictions;
Fig.~\ref{fig:mass-scan} is a two-dimensional scan at these fixed inputs.
The $R$ symmetry protects the small splitting, but does not enforce the required proximity
of the bino, wino and down-type Higgsino masses. These are adjusted to obtain
the transition strength and relic abundance.
The relic calculation follows both possible orderings of the neutral wino and
down-type Higgsino, retaining one solution when the observed abundance is found.
``Relic density not found'' means that varying the wino mass over the tested
range did not yield the observed abundance. It does not exclude a solution
outside that range or with different fixed inputs.

Spectra were generated from SARAH model files
\cite{Staub:2013tta} with SPheno \cite{Porod:2003um,Porod:2011nf} and FlexibleSUSY
\cite{Athron:2014yba,Athron:2017fvs}; relic densities and scattering rates were
calculated with micrOMEGAs \cite{Belanger:2018ccd}, Higgs constraints with HiggsTools
\cite{Bahl:2022igd}, collider limits with SModelS
\cite{Alguero:2021dig,Altakach:2024jwk}, and heavy-Higgs production inputs with SusHi
\cite{Harlander:2012pb,Harlander:2016hcx}.

The relic calculation includes the leading Sommerfeld correction. We estimate
the next-to-leading-order (NLO) shift with a factor $1.035$, the midpoint of the
$2$--$5\%$ correction found for pure winos away from resonances
\cite{Beneke:2020vff}. This is a calibration for the mixed spectrum, not a complete
NLO calculation.
\footnote{Without the wino-specific NLO calibration, the minimal model of
Ref.~\cite{Lee:2026jxl} has a slightly lower relic endpoint. Its present-day
coupled-channel enhancement also tightens the conservative solar bound, leaving
that model's reference point close to the limit.}

Fig.~\ref{fig:mass-scan} shows the dark-matter mass versus (a) the splitting
and (b) the effective mixing strength. The mixing is normalized to one expected
LZ event following Ref.~\cite{Lee:2026jxl}.
Unhatched blue marks spectra passing the adopted tests, including the full solar
uncertainty envelope and finite vacuum searches. Shading interpolates between
calculated points, checked against independent spectra. Only one solution is
displayed when several relic-density solutions are found at the same point.

The solid solar curve uses the nominal calculated-cooling prescription of
Ref.~\cite{Lee:2026jxl}, recalculated for each MRSSM spectrum, rather than the
minimal model's numerical ceiling or its conservative fully thermalized limit.
The dashed curve gives the more restrictive edge from variations in the neutrino
response, present-day Sommerfeld enhancement and electroweak coupling scale,
with the cooled effective volume held fixed. In the diagonally hatched band,
the nominal prediction passes but some variations fail. This is model sensitivity,
not a statistical confidence interval.

For these reference inputs, elastic scattering excludes part of the low-mass
slice (dash-dotted boundary).
At larger masses, compressed charged-wino spectra are cross-hatched because their
collider lifetimes remain unresolved (dotted boundary; Sec.~\ref{sec:pred}); this
is not an exclusion. Solid grey marks cases where the relic density was not
found. Purple stippling marks the region outside the approximate $95\%$ bino-content
selection; its solid purple boundary is the proxy $\st=0.05$.
The thin grey curve marks the approximate switch between selected solutions in
which the neutral wino is heavier or lighter than the down-type Higgsino.
Neither curve is an experimental exclusion.
Spectra satisfying the nominal solar bound occur in both mass orderings.

The dashed and dotted grey curves bound the adopted recoil interval. Light
hatching marks splittings below it, where part of the recoil band falls below
the $55\keV$ threshold and enters the companion's low-energy likelihood.
These splittings are outside our selection and are not classified here.
Thin unsupported edge strips remain white and unclassified.

The blue spectra pass tree-level vacuum tests in the neutral Higgs--adjoint,
$R$-Higgs, stau and stop subspaces; these finite searches do not establish a
complete ultraviolet vacuum or global loop-level stability. Elastic rates use
the physical split states. The solar treatment retains elemental capture, an
orbit and relaxation approximation, scalar isospin and spin-dependent cooling
on hydrogen, rather than complete coupled-channel population evolution.
Exclusions therefore concern the selected spectra on this slice, not the MRSSM
generally.

The constraints shown in Fig.~\ref{fig:mass-scan} depend on the values of
$\tan\beta$, $M_{\rm IR}$ and $\mu_u$ held fixed in the figure.
Varying these parameters allows viable spectra in parts of the plane excluded
for the reference inputs, particularly where elastic scattering is the limiting
constraint.

For comparison with Ref.~\cite{Lee:2026jxl}, we choose an example with $m_\chi=730\GeV$
and $\delta=300\keV$. With the gauge--Higgs boundary condition at
$M_{\rm IR}=10^{14}\GeV$, we evolve the couplings and adjust $M_B^D$, $M_W^D$,
$\mu_d$ and $B_\mu$ at $\tan\beta\simeq31$, keeping the heavy-sector inputs fixed.
The resulting pole spectrum has
\begin{equation}
 \begin{aligned}
 m_\chi&=730\GeV, & \st&\simeq0.02,\\
 \Delta m_{\tilde H_d^0}&\simeq10\GeV,
 &\Delta m_{\tilde W^0}&\simeq11\GeV,
 &\Delta m_{\tilde W^\pm}&\simeq5\GeV.
 \end{aligned}
\label{eq:benchmark-spectrum}
\end{equation}
It gives $\Omega h^2=0.12$ and satisfies the adopted Higgs, flavor, electroweak
and collider constraints, as well as the tests in the specified vacuum subspaces.
Its elastic cross section is $3.5\times10^{-48}\,{\rm cm^2}$, and its nominal
solar-neutrino signal is $0.59$ times the adopted IceCube limit.
The example illustrates the additional states required by the supersymmetric
realization; it does not select a unique mass or splitting.

\section{Low-energy predictions}\label{sec:pred}
We compare the MRSSM realization with the minimal inelastic model of
Ref.~\cite{Lee:2026jxl} before giving its elastic, solar and collider predictions.
The rates below are evaluated at the benchmark of Sec.~\ref{sec:benchmark},
using couplings evolved from the high-scale boundary conditions.

\subsection{Comparison with the minimal model}
The MRSSM realizes the same leading inelastic $Z$ interaction as the minimal model.
The reduced matrix described in Sec.~\ref{sec:embed} explains this when the
wino and up-type Higgsino admixtures are small: $\lambda_d=g'/\sqrt2$ gives equal
mixing in the left- and right-handed components, as does the charge-conjugation
condition of Ref.~\cite{Lee:2026jxl}. All components are retained in the numerical
rates and relic abundance. We set this relation at $M_{\rm IR}$ because it
is one of the coupling relations required by $\mathcal N=2$ supersymmetry.
The low-energy couplings are obtained by running from that scale, so the
relation need not hold exactly at low energy. At fixed
$(m_\chi,\delta,\st)$, the leading $Z$-mediated recoil kinematics and rate are
therefore the same.

The minimal model uses one mixing angle for both chiralities. Before the small
$R$-breaking insertions, the MRSSM has four neutral Dirac states and separate
rotations $N_1$ and $N_2$ defined in Eq.~(\ref{eq:neutralino-rotations})
\cite{Diessner:2014ksa}. Thus $\st$ in Eq.~(\ref{eq:transition-strength}) denotes the physical
transition strength, not a single angle that diagonalizes the neutralino matrix.
The additional up-type Higgsino and wino components, together with the evolution
of the couplings, change this relation and generate a small diagonal axial
current \cite{Benakli:2018vqz}. Both currents are retained in the scattering calculation.

The spectra also differ. The gauge-sized adjoint--Higgs coupling permits a mass
difference of order $10\GeV$ between the Higgsino and bino, rather than the sub-GeV gap of the minimal
model. A percent-level relation between the Dirac and Higgsino masses is still
required. The additional wino supplies coannihilation for the relic abundance,
while the up-type Higgsino and scalar sector alter Higgs-mediated elastic
scattering. The resulting solar cooling and charged-state constraints are not
those of the minimal model, even when the inelastic transition is matched.

The Majorana terms are also different: anomaly mediation gives masses to the
gauginos, but not their adjoint partners, instead of the symmetric singlet terms
assumed in Ref.~\cite{Lee:2026jxl}. The small insertions split the light Dirac state
into two Majorana states whose dominant vector $Z$ interaction is off diagonal.
Eq.~(\ref{eq:measure}) relates their splitting to $F_{\rm IR}$. The warped
stabilization sector generates this small scale; its fields are not additional
entries in the neutralino matrix.

\subsection{The elastic channel}\label{sec:elastic}
The MRSSM's up-type Higgsino pair contributes to elastic scattering in addition to
the down-type pair present in the minimal model. In the Higgs alignment limit,
let $Y_h$ be the neutralino coupling matrix obtained by dividing the
$v_{u,d}$-dependent entries of Eq.~(\ref{eq:mrssm-matrix}) by
$v=\sqrt{v_u^2+v_d^2}$. Using the rotations of Eq.~(\ref{eq:neutralino-rotations})
to express this matrix in the mass basis gives
\begin{equation}
 g_{h\chi\chi}=\left|\bigl(N_1Y_hN_2^T\bigr)_{11}\right|.
\label{eq:ghsusy}
\end{equation}
The down- and up-type contributions interfere destructively at the reference
point. The numerical rates use the physical Majorana eigenstates and retain
scalar mixing and heavy-Higgs exchange. The xenon-equivalent spin-independent
cross section is $\sigma_n=3.5\times10^{-48}\,{\rm cm^2}$ at the reference
point, below the LZ 2024 limit and within the projected XLZD reach
\cite{LZ:2024zvo,XLZD:2024nsu}. This point is not near an elastic blind spot. Nucleon amplitudes
use heavy-quark masses matched to the spectrum's Yukawa couplings; the residual
normalization dependence is a leading-order uncertainty. Spin-dependent scattering
is well below the adopted direct-detection limits, but its contribution to solar
cooling is retained.

\subsection{The solar-neutrino bound}
The captured ground state annihilates almost equally into $W^+W^-$ and $ZZ$;
fermion channels are helicity suppressed. The coupled-channel correction from the
down-type Higgsino pairs gives a present-day rate of
$7.9\times10^{-30}\,{\rm cm^3\,s^{-1}}$ at the reference point. This estimate
includes the leading Sommerfeld effect but not electroweak NLO or bound-state effects.
Related solar-capture analyses for inelastic dark matter are given in
Refs.~\cite{DiMauro:2026dqp,Girmohanta:2026qzo,Nguyen:2026lui,Bose:2026ndd,DiMauro:2026ymt}.
Solar-neutrino limits on elastic interpretations of the LZ event are studied
in Ref.~\cite{Bose:2026szs}.

Captured particles cool efficiently in the nominal orbit and effective-volume
cooling prescription of Ref.~\cite{Lee:2026jxl}. We retain the physical scalar proton and
neutron amplitudes and add spin-dependent scattering on hydrogen. Even when
the axial current is small, this contribution can dominate cooling if the
scalar rate is suppressed. Folding the annihilation spectra with the ten-year
IceCube response gives $\Gamma_{\rm ann}/\Gamma_{\rm lim}=0.59$, with a range of
$0.47$--$0.90$ from the neutrino response and Sommerfeld uncertainties. The reference
point passes the solar bound for this cooling solution, including the alternative
electroweak coupling scale. The cooling calculation uses the same relaxation
approximation as Ref.~\cite{Lee:2026jxl}, rather than a coupled evolution of captured
populations. Inelastic capture and the evolution of the captured phase-space
distribution have been studied in Refs.~\cite{Blennow:2015hzp,Blennow:2018xwu}.
Lighter right-handed sleptons would increase the
$\tau^+\tau^-$ channel; this contribution is included in the solar-neutrino constraint.

The relic initially populates both pseudo-Dirac states. The excited state decays by
photon emission \cite{He:2026hqz} or to neutrinos well before recombination, with
negligible energy release.

\subsection{Other targets and colliders}
At fixed dark-matter mass and splitting, the inelastic target dependence is the same
as in the minimal model of Ref.~\cite{Lee:2026jxl}. Both models have the off-diagonal $Z$
interaction, with minimum incident speed on a nucleus $N$ given by
\begin{equation}
 v_{\min}^{*}=\sqrt{\frac{2\delta}{\mu_{\chi N}}}.
\label{eq:vmin-targets}
\end{equation}
Here $\mu_{\chi N}$ is the dark-matter--nucleus reduced mass. At the reference point,
xenon and tungsten remain accessible, whereas the germanium
and argon thresholds exceed the standard-halo ceiling adopted in Ref.~\cite{Lee:2026jxl}.
The Large-Magellanic-Cloud tail considered in Refs.~\cite{Besla:2019xbx,Bozorgnia:2026emp} remains below the
germanium threshold.
This reproduces the inelastic target hierarchy in Appendix B of Ref.~\cite{Lee:2026jxl}.
The MRSSM differs in its much larger Higgs-mediated elastic signal
(Sec.~\ref{sec:elastic}); the closed inelastic channels do not exclude an elastic
signal in germanium or argon.
Ref.~\cite{Zhang:2026hmg} studies how simulation-informed velocity distributions
affect the inferred inelastic splitting.
Heavy elements also extend the reach of proposed luminous-dark-matter searches
and paleodetectors for inelastic Higgsinos \cite{Graham:2024syw,Graham:2026ivn}.
Ref.~\cite{Graham:2026huo} studies complementary heavy-element searches
and their sensitivity to the high-velocity halo tail.

The charged spectrum also differs from the nearly degenerate doublet of the minimal
model. The MRSSM's down-type charged Higgsino decays promptly, while the additional
charged wino lies about $5\GeV$ above the bino at the reference point. Its decay
length of about $0.16\,{\rm mm}$ gives negligible disappearing-track efficiency
\cite{Gignac:2022fhw}. At
larger masses, some relic-density solutions require a smaller mass difference between the wino and bino, and
the charged wino becomes long lived. We leave spectra with gaps below $3\GeV$
collider-lifetime uncertain, the cross-hatched blue region of Fig.~\ref{fig:mass-scan},
rather than experimentally excluded. Width-rescaling and experimental
efficiency-map tests quantify sensitivity to the lifetime, but do not replace
a calculation of the physical exclusive decay channels. No upper mass boundary
is inferred from these cases. The approximately Dirac gluino
suppresses same-sign dileptons, while the small $R$ breaking suppresses electric
dipole moments and flavor violation \cite{Kribs:2007ac,Kribs:2010md}.

\section{Conclusion}\label{sec:concl}
In this paper, we have shown that the MRSSM can realize the mass, splitting and
suppressed $Z$ coupling required by an inelastic interpretation of the LZ
$248\keV$ candidate. Its $U(1)_R$ symmetry allows the bino and singlet adjoint
to form a Dirac fermion, while small Majorana terms set the splitting.
Mixing with the Higgsinos supplies the inelastic $Z$ interaction, and the
high-scale $\mathcal N=2$ coupling relations help suppress the diagonal axial
current. Wino coannihilation gives the observed relic abundance. With couplings
evolved from the high-scale boundary, we find spectra over a broad mass range
for the chosen fixed inputs that satisfy the relic-density, Higgs, flavor,
electroweak and collider tests, as well as the specified vacuum-subspace tests.
The small splitting is technically natural, controlled by symmetry breaking
rather than a cancellation between independent mass parameters. In contrast,
the MSSM Higgsino limit fails the adopted solar bound, while the gaugino alternative
requires an unprotected mass degeneracy.

When anomaly mediation dominates, the inelastic splitting probes the local
compensator scale, which is tens of MeV for a splitting of a few hundred keV.
Warping allows this scale to be small while the gravitino remains heavier than
the dark matter. In the matched warped realization, a
Goldberger--Wise hypermultiplet fixes the radius and a loop correction to the
no-scale K\"ahler potential generates the suppressed infrared compensator.
The Dirac-mass sector does not appreciably shift this vacuum. Gravitational loops
are included in the local compensator rather than added as an independent
Majorana source. The result assumes sequestering and a reheating history that
preserves the thermal relic. A bulk-condensate realization offers an alternative,
but its connection to a microscopic stabilization sector remains to be established.
A common gravitino-mass convention and consistent loop-level boundary matching
are also needed.

The construction offers several complementary tests. Further high-energy xenon
recoils would constrain the mass, splitting and transition strength, while
comparison with other target nuclei would test the inelastic threshold pattern.
Elastic scattering probes the Higgsino components that also enter solar cooling,
and solar neutrinos test capture and annihilation of the same dark-matter states.
At the reference benchmark, the elastic rate lies within the projected XLZD
reach and the solar-neutrino signal is below but near the IceCube limit.
Charged-state searches probe the nearby wino required for coannihilation.
Improved solar-cooling and charged-state lifetime predictions would sharpen
these comparisons. Together, these measurements can help distinguish the MRSSM
from realizations with the same leading inelastic interaction but different
elastic rates and coannihilating partners, and test its proposed connection to
supersymmetry breaking.

\acknowledgments
We thank Lisa Randall and Huangyu Xiao for discussions.
SJL and TY are supported by the Samsung Science and Technology Foundation under
Project Number SSTF-BA2201-06; TY is also supported by the GRASP Initiative at Harvard University.

\appendix
\section{Five-dimensional warped construction}\label{sec:seq}
This appendix specifies the bulk and ultraviolet boundary sectors of
Eq.~(\ref{eq:matched-5d-action}) and collects the checks of the two stabilization
models in Fig.~\ref{fig:matched-mediation}. We use the direct five-dimensional auxiliary fields
of Ref.~\cite{Nee:2025nmi}; quantities that depend on the physical gravitino
normalization should be read as estimates.
The Goldberger--Wise and bulk-condensate samples were evaluated at
$F_{\rm IR}=30.6\MeV$ and $31.6\MeV$, respectively. These inputs lie within
a few percent of the $31\MeV$ scale used in the main text, smaller than the
normalization uncertainty in Fig.~\ref{fig:matched-mediation}.

\subsection{Bulk fields and the ultraviolet boundary}\label{sec:bulk-fields}
The off-shell theory contains the gravitational Weyl multiplet, a graviphoton
multiplet, the compensator hypermultiplet and the physical Goldberger--Wise
hypermultiplet \cite{Abe:2004ar,PaccettiCorreia:2006rpo,Nee:2025nmi}.
The Weyl multiplet includes the metric and gravitini; its remaining connections
and auxiliary fields are gauge fixed or eliminated. The graviphoton multiplet
similarly includes a scalar, fermions and an auxiliary triplet. Its scalar is
fixed by the conformal gauge, leaving the superfields $V$ and $\Sigma$ used in
the main text. The chiral fields $C_+,\Phi_+,\Sigma$ are even under the orbifold
reflection, whereas $C_-,\Phi_-,V$ are odd. Only the physical hypermultiplet
adds bulk matter: $\Phi_\pm$ contain complex scalars $\varphi_\pm$, hyperinos
and auxiliary fields $F_{\Phi_\pm}$.

For the bosonic vacuum calculation, the two bulk densities in
Eq.~(\ref{eq:matched-5d-action}) can be written together as
\cite{Abe:2004ar,Nee:2025nmi}
\begin{align}
 \mathcal L_{\rm SUGRA}^{(5)}+\mathcal L_{\rm hyp}^{(5)}
 ={}&\mathcal L_{\rm vec}+\mathcal L_K+\mathcal L_W,\nonumber\\
 \mathcal L_{\rm vec}={}&-\frac32\left[\int d^2\theta\,
 \Sigma\mathcal W^\alpha\mathcal W_\alpha+{\rm h.c.}\right]\nonumber\\
 &+\frac18\left[\int d^2\theta\,\overline D^2
 \left(VD^\alpha\partial_yV-D^\alpha V\,\partial_yV\right)
 \mathcal W_\alpha+{\rm h.c.}\right],\nonumber\\
 \mathcal L_K={}&-3e^{2\sigma}\int d^4\theta\,\mathcal V_\Sigma
 \Bigl[C^\dagger_a(e^{-2g_c\sigma_3 V})^a{}_b C^b\nonumber\\
 &\hspace{7em}-\Phi^\dagger_a(e^{-2g_h\sigma_3 V})^a{}_b\Phi^b\Bigr]^{2/3},\nonumber\\
 \mathcal L_W={}&e^{3\sigma}\Bigl[\int d^2\theta\,
 \Phi_a(\partial_y+2g_h\Sigma\sigma_3)^a{}_b\Phi^b\nonumber\\
 &\hspace{4em}-\int d^2\theta\,
 C_a(\partial_y+2g_c\Sigma\sigma_3)^a{}_b C^b
 +{\rm h.c.}\Bigr].
\label{eq:bulk-superfield-densities}
\end{align}
Here $a,b=+,-$ label the two chiral components of each hypermultiplet,
with indices raised and lowered by the antisymmetric tensor, and
$\sigma_3=\operatorname{diag}(1,-1)$.
The field strength $\mathcal W_\alpha=-\overline D^2D_\alpha V/4$ belongs to
the graviphoton, not an MRSSM gauge field.
On $0<y<\pi$, the gauging couplings satisfy $2g_cM_5=3k$ and
$g_h=-g_c(1+2\epsilon/3)$; their continuation across the orbifold is odd.
The conformal gauge sets the scalar component of $C_-$ to zero and fixes
$|C_+|^2=M_5^3+|\varphi_+|^2+|\varphi_-|^2$.
Neglecting the small hypermultiplet backreaction gives $C_+=M_5^{3/2}$.
The source, loop and constant-superpotential terms displayed in
Eq.~(\ref{eq:matched-5d-action}) are added to Eq.~(\ref{eq:bulk-superfield-densities}).
The fermionic completion is given in Refs.~\cite{Abe:2004ar,PaccettiCorreia:2006rpo}.

The ultraviolet multiplet has components
$Q_0=q_0+\sqrt2\theta\psi_0+\theta^2F_{Q_0}$: a hidden scalar, a fermion and an
auxiliary field. Its boundary action has the standard compensator form
\begin{equation}
 \mathcal L_{\rm uplift}
 =\int d^4\theta\,C_0^\dagger C_0\,
 \Omega_{\rm hid}(Q_0^\dagger,Q_0)
 +\left[\int d^2\theta\,C_0^3W_{\rm hid}(Q_0)+{\rm h.c.}\right].
\label{eq:uv-uplift-action}
\end{equation}
Here $C_0$ is the dimensionless ultraviolet boundary compensator, proportional
to $C_+(0)^{2/3}$ and normalized to unit scalar expectation value.
The real function $\Omega_{\rm hid}$ specifies the hidden-sector kinetic
couplings, and $W_{\rm hid}$ is its superpotential. Direct couplings to the
infrared MRSSM are absent by the assumed sequestering.
We require the hidden action and its vacuum to preserve $R$-parity.
We use a fixed hidden
vacuum with a canonically normalized auxiliary field, so the leading positive
contribution is $V_{\rm uplift}\simeq|F_{Q_0}|^2$. Its magnitude is adjusted
to make the total four-dimensional vacuum energy approximately zero, including
the infrared $D'$ contribution. The microscopic functions
$\Omega_{\rm hid},W_{\rm hid}$ and the dynamics fixing $q_0$ are not specified;
the radion and mediation checks use the effective boundary energy source.

\subsection{Component reduction of the Goldberger--Wise realization}\label{sec:gw-reduction}
The Goldberger--Wise hypermultiplet contains an even field $\Phi_+$ and an odd field
$\Phi_-$. The component equation following from the bulk hypermultiplet action and
the $J_0,J_\pi$ terms displayed in Eq.~(\ref{eq:matched-5d-action}) is
\begin{align}
 F_{\Phi_+}^\dagger={}&-\frac{e^\sigma}{r}
 \left(\partial_y+\frac32\dot\sigma-g_hM_5r\right)\varphi_-
 +\frac{2e^\sigma M_5^{3/2}}{r}
 \left[\delta(y)J_0-\delta(y-\pi)J_\pi\right],\nonumber\\
 \widetilde\varphi_-(0)={}&M_5^{3/2}J_0,\qquad
 \widetilde\varphi_-(\pi)=M_5^{3/2}J_\pi,\qquad
 \varphi_-=\Theta(y)\widetilde\varphi_-.
\label{eq:gw-boundary-equations}
\end{align}
The boundary conditions in the second line cancel the delta functions in the first.
In the bulk, $F_{\Phi_+}=0$ then gives
\begin{equation}
 \widetilde\varphi_-(y)=M_5^{3/2}J_0
 \exp\!\left[-\frac32\sigma(y)+rM_5g_hy\right].
\label{eq:gw-flat-profile}
\end{equation}
Using $2g_cM_5=3k$ and
$g_h=-g_c(1+2\epsilon/3)$, the infrared boundary condition becomes
$\rho_0^\epsilon=J_\pi/J_0$. For a general radius, the bulk solution inserted into
the positive $F$-term integral gives
\begin{align}
 V_{\rm GW}={}&\frac{1}{r}\int_0^\pi dy\,e^{4\sigma}
 \left|\partial_y\varphi_--M_5r(g_c+g_h)\varphi_-\right|^2\nonumber\\
 ={}&\frac{2M_5^4(g_c-3g_h)
 \left|J_\pi-J_0e^{\pi M_5r(g_c+g_h)}\right|^2}
 {3\left[e^{8\pi g_cM_5r/3}-e^{2\pi M_5r(g_c+g_h)}\right]}.
\label{eq:gw-profile-potential}
\end{align}
Substituting the same definitions of $g_c$ and $g_h$, writing
$\rho=e^{-k\pi r}$, and expanding at small $\rho$ gives
Eq.~(\ref{eq:gw-potential}).

At $\beta=0$, the no-scale solution has $F_{C_+}=0$. The two terms
$\Delta\mathcal L_\beta$ and $\mathcal L_{W_{\rm bulk}}$ perturb the radion and
compensator auxiliary equations, giving Eq.~(\ref{eq:matched-auxiliaries}) at first
order in $\beta$. Dividing its second relation by $C_+=M_5^{3/2}$, evaluating
$e^\sigma=\rho_0$ at the infrared boundary, and using
$m_{3/2}=M_5|W_{\rm bulk}|$ gives
\begin{equation}
 \left|\frac{F_{C_+}}{C_+}\right|_{y=\pi}
 =\frac{2\beta\rho_0|W_{\rm bulk}|}{M_5^3r^4}
 =\frac{2\beta}{(M_5r)^4}\rho_0m_{3/2}
 =c_\beta\rho_0m_{3/2}.
\label{eq:appendix-local-scale}
\end{equation}
This is Eq.~(\ref{eq:matched-loop-scale}); Eqs.~(\ref{eq:dtt}) and
(\ref{eq:measure}) then perform the anomaly-mediation and light-state projection
steps in the main text.

We use the direct five-dimensional normalization in
Eq.~(\ref{eq:appendix-local-scale}). Combining two alternative normalizations quoted
in Ref.~\cite{Nee:2025nmi} changes $c_\beta$ by a factor $\sqrt3$; this spread is
included in Fig.~\ref{fig:matched-mediation}. The boundary compensator has
conformal weight one, whereas $C_+$ has weight $3/2$; hence
$C_{\rm bdy}=C_+^{2/3}$. Expanding this relation in superspace gives
\begin{equation*}
 \frac{F_{C_{\rm bdy}}}{C_{\rm bdy}}
 =\frac23\frac{F_{C_+}}{C_+}.
\end{equation*}
These auxiliary ratios cannot be identified. Eq.~(\ref{eq:dtt}) uses the direct
five-dimensional boundary mass operator of Ref.~\cite{Nee:2025nmi}, expressed in
terms of $F_{C_+}/C_+$, and $F_{\rm IR}$ retains that normalization throughout.
For the condensate alternative, the matched four-dimensional constant
term instead gives $m_{3/2}=M_5|W_{\rm bulk}|/4$, producing the factor-four band
shown there.

\subsection{Parameter ranges for the two stabilization models}

\paragraph{Goldberger--Wise stabilization with loop corrections.}
At the supersymmetric minimum of Eq.~(\ref{eq:gw-potential}), the canonically normalized
radion mass is
\begin{equation}
 m_\rho=\epsilon J_0 k\rho_0^{1+\epsilon}
 \sqrt{\frac{4+2\epsilon}{3}}.
\label{eq:gw-radion-mass}
\end{equation}
For the numerical examples, we fix $F_{\rm IR}=30.6\MeV$ through
Eq.~(\ref{eq:matched-loop-scale}) and use
\begin{equation*}
 \epsilon=J_0=0.1,\qquad
 \beta=\frac{1\text{--}4}{16\pi^2},\qquad
 k/M_5=0.3\text{--}1.
\end{equation*}
The perturbative condition of Ref.~\cite{Nee:2025nmi} is satisfied for these inputs.
The matching permits a broad range of scales: $m_{3/2}=10^4$--$10^7\GeV$
corresponds to roughly $M_{\rm IR}=3\times10^{13}$--$10^{16}\GeV$.

The required ultraviolet uplift is small, with $|F_{Q_0}|$ at most $2\%$ of
$\sqrt3m_{3/2}\MP$. The residual loop-generated vacuum energy only weakly shifts
the stabilized radius: the relative change is below $0.17\%$ for ultraviolet
uplift and below $1.8\%$ for a purely infrared uplift.

\paragraph{Bulk gaugino condensation.}
Let $p=\alpha W_\lambda$ and $t=\alpha M_5r=x\ln(1/\rho)$. Direct differentiation of
the displayed five-dimensional potential of Ref.~\cite{Nee:2025nmi} gives
\begin{equation}
 p\rho^x=\frac{3|W_{\rm bulk}|}{2+6/t+3/t^2}
 =\frac{c_1}{4}|W_{\rm bulk}|.
\label{eq:condensate-stationary}
\end{equation}
The printed stationary condition and the four-dimensional matching discussion of
Ref.~\cite{Nee:2025nmi} quote different coefficients; we use the direct
differentiation in Eq.~(\ref{eq:condensate-stationary}), which agrees asymptotically
with their printed minimum. The infrared scale is then
Eq.~(\ref{eq:matched-condensate-scale}), and
\begin{equation}
 m_\rho\simeq\frac{2x\,m_{3/2}}{\rho_0}.
\label{eq:condensate-radion-mass}
\end{equation}
Five of the twelve sampled points obey both $m_{3/2}>730\GeV$ and
$m_\rho<M_{\rm IR}$. They span the range quoted in the main text, but none reaches the
$60\TeV$ reference for a standard thermal gravitino history.

\subsection{Gravitational loops and the boundary gaugino}\label{sec:gravity-loop}
The infrared-brane gauge kinetic term is holomorphic and Weyl neutral. In the
sequestered brane action, the bulk contribution to its Majorana mass enters through
the local anomalous operator in Eq.~(\ref{eq:dtt})
\cite{Randall:1998uk,Luty:2000ec,Luty:2001jh,Luty:2001zv,Nee:2025nmi}. The one-loop effect of the bulk
no-scale auxiliary field is precisely the $\beta$ correction that generates
$F_{C_+}(\pi)$ in Eq.~(\ref{eq:matched-auxiliaries}), not an additional mass.

A loop sourced by the ultraviolet uplift is further suppressed. Relative to the term
already used in Eq.~(\ref{eq:measure}), its size is parametrically
\begin{equation}
 \frac{\Delta m_B^{\rm UV}}{m_B^{\rm AM}}
 \sim \frac{\rho_0}{16\pi^2c_\beta}
 \frac{F_{Q_0}}{\sqrt3m_{3/2}\MP}
 \simeq9\times10^{-5}
 \quad\text{at }m_{3/2}=10^5\GeV .
\label{eq:uplift-gaugino-ratio}
\end{equation}
Flat-space brane-to-brane gaugino masses are not generated at one loop
\cite{Gherghetta:2001sa,Rattazzi:2003rj}, while the
known warped calculation produces K\"ahler and scalar corrections rather than a
separate gaugino term \cite{Gregoire:2004nn}. The dimensional estimate imported from
four-dimensional cutoff calculations \cite{Barbieri:1983af,Luty:2002ff} consequently does not define
an independent Majorana mass for the infrared-brane bino in this construction.

\subsection{Backreaction from the Dirac-mass sector}\label{sec:radion}
The earlier theory with independent boundary superpotentials \cite{Luty:2000ec,Luty:2002ff}
does not reproduce the matched five-dimensional potential \cite{Nee:2025nmi}.
Its sensitivity to the $D'$ energy therefore does not apply to the stabilization
models evaluated here.

Across the Goldberger--Wise parameter range, the largest linearized fractional
shift of the warp factor is $7\times10^{-12}$. In the condensate sample the largest shift
is $1.2\times10^{-6}$, including a loop-suppressed supersoft coefficient
$\kappa_D=1/411$. In these checks, $D'=(1\TeV)M_{\rm IR}/\kappa_D$.
The Goldberger--Wise uplift range quoted above uses $\kappa_D=1$; for a smaller
coefficient, the $D'$ energy can contribute substantially to the vacuum-energy
budget even while the radius remains stable. These hierarchies make a dedicated
radion stabilizer for the $D'$ sector unnecessary. What remains open is the physical
gravitino normalization in a common fermion convention, the factor-$\sqrt3$ difference
among the quoted formulas for the Goldberger--Wise realization, the loop-level boundary matching, and the
microscopic map from the bulk condensate to the zero-mode $SU(2)$ stabilizer of
Ref.~\cite{Luty:2002ff} with ultraviolet-brane flavours. The standard thermal history also
requires the reheating bound discussed below.

\subsection{Thermal gravitinos}\label{sec:gravitino-history}
With the width $\Gamma_{3/2}=0.17m_{3/2}^3/\MP^2$, gravitinos with masses between
$10^4$ and $10^7\GeV$ decay after bino freeze-out. Each decay therefore adds a
bino to the thermal relic. For the $730\GeV$ reference point, limiting this component
to ten percent of the observed abundance requires
\begin{equation*}
 Y_{3/2}\lesssim6\times10^{-14},\qquad T_R\lesssim3\times10^8\GeV,
\end{equation*}
using the standard thermal yield \cite{Bolz:2000fu,Pradler:2006qh}.
The gravitino never dominates the energy density, and the Goldberger--Wise radion is
heavy, so this bound preserves the thermal relic calculation. The
$60\TeV$ nucleosynthesis reference applies only to the assumed thermal gravitino
history \cite{Gherghetta:1999sw,Kawasaki:2008qe}.

\bibliographystyle{JHEP}
\bibliography{refs}
\end{document}